\documentstyle[psfig]{mn}

\newif\ifAMStwofonts

\ifoldfss
  \ifCUPmtlplainloaded \else
    \NewTextAlphabet{textbfit} {cmbxti10} {}
    \NewTextAlphabet{textbfss} {cmssbx10} {}
    \NewMathAlphabet{mathbfit} {cmbxti10} {} 
    \NewMathAlphabet{mathbfss} {cmssbx10} {} 
  \fi
  \ifAMStwofonts
    \ifCUPmtlplainloaded \else
      \NewSymbolFont{upmath} {eurm10}
      \NewSymbolFont{AMSa} {msam10}
      \NewMathSymbol{\upi}     {0}{upmath}{19}
      \NewMathSymbol{\umu}     {0}{upmath}{16}
      \NewMathSymbol{\upartial}{0}{upmath}{40}
      \NewMathSymbol{\leqslant}{3}{AMSa}{36}
      \NewMathSymbol{\geqslant}{3}{AMSa}{3E}

      \let\leq=\leqslant \let\le=\leqslant
       
    \fi
  \fi
\fi 

\ifnfssone
  \newmathalphabet{\mathit}
  \addtoversion{normal}{\mathit}{cmr}{m}{it}
  \addtoversion{bold}{\mathit}{cmr}{bx}{it}
  \newmathalphabet{\mathbfit} 
  \addtoversion{normal}{\mathbfit}{cmr}{bx}{it}
  \addtoversion{bold}{\mathbfit}{cmr}{bx}{it}
  \newmathalphabet{\mathbfss} 
  \addtoversion{normal}{\mathbfss}{cmss}{bx}{n}
  \addtoversion{bold}{\mathbfss}{cmss}{bx}{n}
  \ifAMStwofonts
    \ifCUPmtlplainloaded \else
      \UseAMStwoboldmath
      \makeatletter
      \new@mathgroup\upmath@group
      \define@mathgroup\mv@normal\upmath@group{eur}{m}{n}
      \define@mathgroup\mv@bold\upmath@group{eur}{b}{n}
      \edef\UPM{\hexnumber\upmath@group}
      \new@mathgroup\amsa@group
      \define@mathgroup\mv@normal\amsa@group{msa}{m}{n}
      \define@mathgroup\mv@bold\amsa@group{msa}{m}{n}
      \edef\AMSa{\hexnumber\amsa@group}
      \makeatother
      \mathchardef\upi="0\UPM19
      \mathchardef\umu="0\UPM16
      \mathchardef\upartial="0\UPM40
      \mathchardef\leqslant="3\AMSa36
      \mathchardef\geqslant="3\AMSa3E

      \let\leq=\leqslant \let\le=\leqslant

    \fi
  \fi
\fi 

\ifnfsstwo
  \DeclareMathAlphabet{\mathbfit}{OT1}{cmr}{bx}{it}
  \SetMathAlphabet\mathbfit{bold}{OT1}{cmr}{bx}{it}
  \DeclareMathAlphabet{\mathbfss}{OT1}{cmss}{bx}{n}
  \SetMathAlphabet\mathbfss{bold}{OT1}{cmss}{bx}{n}
  \ifAMStwofonts
    \ifCUPmtlplainloaded \else
      \DeclareSymbolFont{UPM}{U}{eur}{m}{n}
      \SetSymbolFont{UPM}{bold}{U}{eur}{b}{n}
      \DeclareSymbolFont{AMSa}{U}{msa}{m}{n}
      \DeclareMathSymbol{\upi}{0}{UPM}{"19}
      \DeclareMathSymbol{\umu}{0}{UPM}{"16}
      \DeclareMathSymbol{\upartial}{0}{UPM}{"40}
      \DeclareMathSymbol{\leqslant}{3}{AMSa}{"36}
      \DeclareMathSymbol{\geqslant}{3}{AMSa}{"3E}

      \let\leq=\leqslant \let\le=\leqslant

    \fi
  \fi
\fi 

\ifCUPmtlplainloaded \else
  \ifAMStwofonts \else 
    \def\upi{\pi}
    \def\umu{\mu}
    \def\upartial{\partial}
  \fi
\fi

\def\ltsima{$\; \buildrel < \over \sim \;$}
\def\gtsima{$\; \buildrel > \over \sim \;$}

\title[Extinction close to the Galactic centre]
{Extinction within 10$^{\circ}$ of the Galactic centre using 2MASS}
\author[C. Dutra et al.]
  {C.~M.~Dutra,$^1$
  B.~X.~Santiago,$^2$ E.~L~D.~Bica,$^2$ B. Barbuy$^1$ \\
  $^1$Universidade de S\~ao Paulo, Instituto de Astronomia, Geof\'\i sica e Ci\^encias atmosf\'ericas, CP 3386, S\~ao Paulo 01060-970, SP, Brazil\\
  $^2$Universidade Federal do Rio Grande do Sul, Instituto de F\'\i sica, 91501-970 Porto Alegre, RS Brasil}

\begin{document}
\maketitle
\label{firstpage}
\begin{abstract}  
We extract {\it J} and {\it K$_s$} magnitudes from the 2MASS Point Source 
Catalog for approximately 6$\times 10^6$ stars with $8 \le {\it K_s} \le 13$ 
in order to build an {\it$A_K$} extinction map within 10$^{\circ}$ of the 
Galactic centre. The extinction was determined by fitting the upper giant 
branch of ({\it K$_s$, J-K$_s$}) colour-magnitude diagrams to a dereddened 
upper giant branch mean locus built from previously studied Bulge fields. The 
extinction values vary from {\it$A_K$}=0.05 in the edges of the map up to 
{\it$A_K$}=3.2 close to the Galactic centre. The 2MASS extinction map was 
compared to that recently derived from DENIS data. Both maps agree very well 
up to {\it$A_K$}=1.0. Above this limit, the comparison is affected
by increased internal errors in both extinction determination methods. 
The 2MASS extinction values were also compared 
to those obtained from dust emission in the far infrared using DIRBE/IRAS.
Several systematic effects likely to bias this comparison were addressed,
including the presence of dust on the background of the bulk
of 2MASS stars used in the extinction determination.
For the region with $3^{\circ}<|{\it b}|<5^{\circ}$, where the dust 
contribution on the far side of the Galaxy is $\approx$ 5 \%, the two 
extinction determinations correlate well, but the 
dust emission {\it$A_K$} values are systematically higher than those from 
2MASS. A calibration correction factor of 76\% for the DIRBE/IRAS dust
emission extinction is needed to eliminate this systematic effect.
Similar comparisons were also carried out for the 
$1^{\circ}<|{\it b}|<3^{\circ}$ and $|{\it b}| < 0.5^{\circ}$ strips, 
revealing an increasing complexity in the relation between the two extinction 
values. Discrepancies are explained in terms of the
calibration factor, increasing background dust contribution, 
temperature effects influencing 
the dust emission extinction and limitations 
in the 2MASS extinction determination in very high extinction regions 
($|{\it b}| < 0.5^{\circ}$). An asymmetry relative to the Galactic
plane is observed in the dust maps, roughly in the sense that
{\it$A_K$} values are 60\% smaller in the south than in the north for 
$1^{\circ}<|{\it b}|<5^{\circ}$. This asymmetry is due to the
presence of foreground dust clouds mostly in the northern region 
of the Bulge.
\end{abstract}


\section{Introduction}
The high extinction and its patchy distribution in the Galactic centre 
region have been a constant problem to the study of the 
properties of the Bulge stellar population. Several efforts were carried 
out to investigate the extinction distribution close to the Galactic 
centre. For example, Catchpole et al. (1990) studied the distribution 
of stars in the central $1^{\circ} \times  2^{\circ}$ of the Galaxy by 
means of {\it J}, {\it H} and {\it K} colour-magnitude diagrams (CMDs). They 
derived visual absorptions in the range  $7 < {\it A_V} < 30$. Frogel 
et al. (1999, hereafter FTK99) obtained extinction values varying from 
{\it$A_V$}=2.41 up to {\it$A_V$}=19.20 
for 11 Bulge fields with $|{\it b}|<4^{\circ}$, close to the Galactic centre. 
Stanek (1996) derived a mean extinction of ${\it<A_V>}=1.54$ for Baade's 
Window (Baade 1963).
As a consequence, the investigations about the Bulge stellar content based 
on optical data were restricted for a long time to 
lower-extinction regions such as Baade's Window. 

The advent of near infrared surveys such as the Two Micron All Sky Survey 
(2MASS, Skrutskie et al. 1997) and the Deep NIR Southern Sky Survey 
(DENIS, Epchtein et al. 1997) has provided fundamental tools to study 
the stellar population (Unavane et al. 1998) and extinction (Schultheis et al.
 1999) in the inner Bulge. Recently, Dutra et al. (2002, hereafter 
Paper I) confirmed the existence of two new Bulge windows, W0.2-2.1 and 
W359.4-3.1, closer to the Galactic centre than Baade's Window; they
used the {\it JK$_s$} photometry
from the 2MASS survey archive to map the extinction distribution within 
1$^{\circ}$ towards these windows. The reddening distribution
in the Bulge area is also fundamental to understand the spatial
distribution of globular clusters and their relation to the Bulge field 
stellar population itself. Barbuy et al. (1998) discussed the globular 
clusters projected within 5$^{\circ}$ of the nucleus and
which appear to be related to the Bulge, while
Barbuy et al. (1999) discussed the properties of those found in the
area covered by the present study.

\begin{figure}
\begin{center}
\centerline{\psfig{file=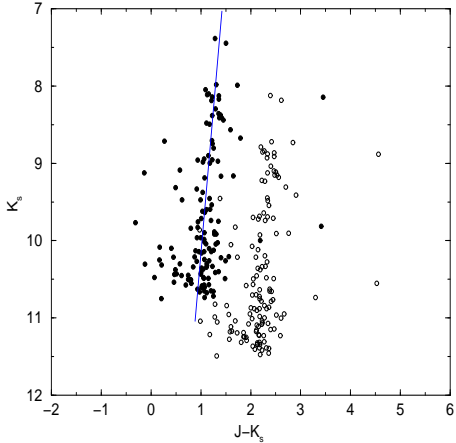,height=6cm,width=6cm,angle=0}}
\end{center}
\caption{Extinction-corrected (filled circles) and observed (open circles) 
CMDs for a cell at $\ell=1.93^{\circ}$, ${\it b}=0.93^{\circ}$. The straight 
line represents the reference upper giant branch (Eq. 1).}
\end{figure}

In this work we use the 2MASS {\it J} (1.25$\mu$m) and {\it$K_s$} (2.17$\mu$m) 
photometric data to build ({\it K$_s$, J-K$_s$}) 
CMDs of Bulge fields in order to map out the interstellar extinction 
in the central 10$^{\circ}$ of the Galaxy. The extinction was derived by 
means of the upper giant branch fitting method, 
similar to that described in Paper I. In Sect. 2 we revise this method 
and describe in detail the way we build the extinction map. 
In Sect. 3 we compare the present 
results with those from  DENIS. In Sect. 4 we analyze the 2MASS and 
DIRBE/IRAS extinction maps, making use of simple models for the optical 
depth in different directions in the Galaxy,
and discuss the asymmetries between observations in the northern and 
southern Galactic hemispheres. Finally, the concluding remarks are 
given in Sect. 5.

\section[]{Building the extinction map with 2MASS data}

In Paper I we have built an extinction map in regions of low-extinction 
by means of upper giant branch fitting. We adopted an upper giant branch 
template from a composite CMD using 7 Bulge fields from FTK99. Using 
this template we managed to reproduce the mean extinction 
in Baade's and Sgr I windows. Metallicity effects were not considered 
in Paper I, since Ramirez et al. (2000), from a spectroscopic study of 
central M giant stars, pointed out that there is no 
evidence for metallicity gradient within the inner Bulge ({\it R} $<$ 560 pc).
Besides, FTK99 concluded that the amplitude of metallicity variations in 
the inner Bulge implies very small giant branch slope changes. Therefore, 
we assume that the 
metallicity variations in the inner Bulge do not
affect significantly the extinction estimates and use as reference 
the same upper giant branch adopted in  Paper I:

\begin{equation}
(K_s)_0 = -7.81 (J-K_s)_0 + 17.83
\end{equation}

The equation above describes appropriately the upper giant 
branch locus for stars with 8 $\le {\it K_0} \le $12.5. 

\noindent We carried out {\it JK$_s$} photometric extractions of stars 
in the 2MASS Point Source Catalog available in the Web Interface 
{\it http://irsa.ipac.caltech.edu/applications/Gator/} for 61 fields with
radius {\it r}=1$^{\circ}$ each. 
For most fields, we extracted stars in the range 8 $\le {\it K_s} \le $ 
11.5, there being typically 60,000 such stars in each. 
For the fields with $|{\it b}| <$ 1.5$^{\circ}$, we expected 
very high extinction (Schultheis et al. 1999); thus we extracted fainter stars,
down to {\it$K_s$}=13. This extra 1.5 mag allowed us to determine the
best magnitude range for upper giant branch fitting, taking into account
factors such as the increasing photometric errors and contamination by disk 
stars with fainter magnitude limits, and the decreasing observed 
extent of the upper giant branch in heavily reddened inner Bulge fields. 
This issue is discussed in Sect. 2.1.
In these low-latitude regions the deeper 2MASS extractions led to a
considerably larger number of stars per field, $\approx$ 200,000.
Considering all fields in the area, we extracted {\it J} and {\it$K_s$} 
magnitudes for approximately 6$\times 10^6$ stars.

\begin{figure}
\begin{center}
\centerline{\psfig{file=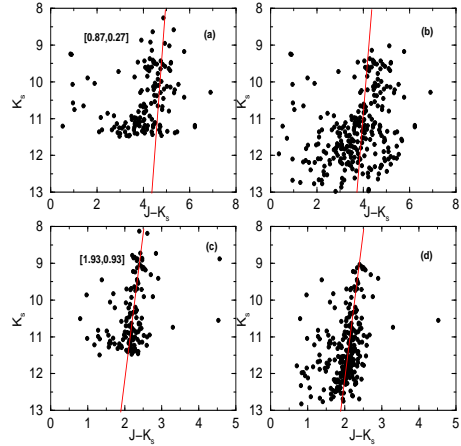,height=6cm,width=6cm,angle=0}}
\end{center}
\caption{Analysis of limiting magnitude for fields with $|{\it b}|<1.5^{\circ}$: CMDs for stars with (a) 8 $\le {\it K_s} \le $ 11.5 and 
(b) 9 $\le {\it K_s} \le $ 13 within a cell 
at $\ell=0.87^{\circ}$, ${\it b}=0.27^{\circ}$; (c) and (d) same {\it$K_s$} intervals 
as in (a) and (b), respectively, for stars within a cell 
at $\ell=1.93^{\circ}$, ${\it b}=0.93^{\circ}$. The straight line represents 
the reddened reference upper giant branch. The derived extinction values in panels (a) and (b) are $A_K$= 1.88 and 1.83, respectively. For (c) and (d) $A_K$= 0.74.}
\end{figure}

\begin{figure*}
\begin{center}
\centerline{\psfig{file=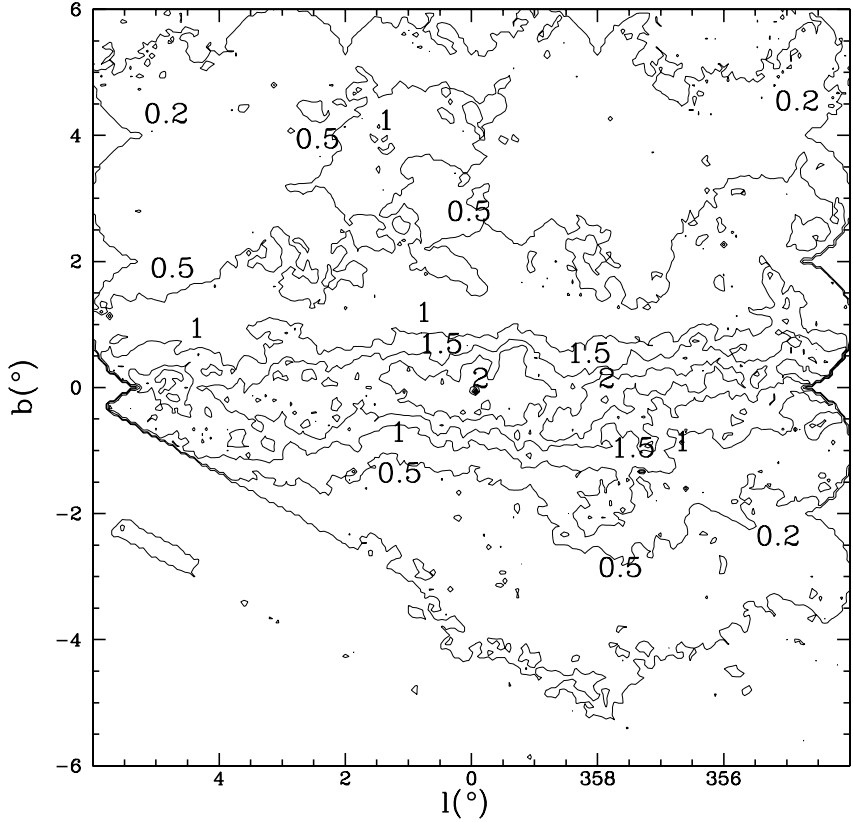,height=12cm,width=12cm,angle=0}}
\end{center}
\caption{{\it$A_{K,2MASS}$} extinction map within 10$^{\circ} \times 10^{\circ}$ 
of the Galactic centre. The contours represent 
the following extinction values: {\it$A_K$} = 0.2, 0.5, 1.0, 1.5 and 2.0.}
\end{figure*} 

In order to map out the extinction in the 10$^{\circ}$ Galactic central 
region, we defined 961 cells with 4$^{\prime} \times 4^{\prime}$ in size  
in each one of 61 extracted fields. The extinction was determined 
assuming that the upper giant branch defined by the stars in a cell has 
the same slope as that adopted as reference (Equation 1). We then
calculated from the ({\it K, J-K$_s$}) values
of each star the shift along the reddening vector necessary to make it 
fall onto the reference upper giant branch. 
We used the relations ${\it A_{K_s}} = 0.670 {\it E(J-K_s)}$ and 
$\frac{{\it A_K}}{{\it A_{K_s}}}=0.95$ (Paper I) to derive the {\it$A_K$} 
values. The {\it$A_K$} adopted for each cell was taken to be the median of 
the distribution of {\it$A_K$} values for the individual stars in it; 
an iterative 2-$\sigma$ clipping was applied to the distribution 
of {\it$A_K$} values in order to eliminate contamination from foreground 
stars. Fig. 1 shows the method applied to a cell located 
at $\ell=1.93^{\circ}$, ${\it b}=0.93^{\circ}$; the stars in the observed CMD
(open circles) have a median extinction of ${\it A_K}=0.74$ 
with respect to the reference upper giant branch.
The extinction-corrected CMD based on this value is also shown (filled circles)
in the figure. Note that the extinction-corrected CMD
is well fitted by the reference upper giant branch (straight line), 
indicating that the determined median extinction is
representative of most stars within the cell's area. 

\subsection{The upper giant branch fitting in heavily reddened fields}

We now try to assess the existence of systematic effects on the extinction
determination just outlined. Pushing the limits used in our fitting method
towards fainter magnitudes is desirable only if it 
increases the magnitude range where the upper giant branch is clearly defined
and can be fitted.
Several factors may prevent fainter magnitudes from being useful in this
way. The first is obviously the increasingly large photometric errors.
The 2MASS {\it JK$_s$} nominal errors are smaller than 0.04 for $J$ \ltsima
$15$ and $K_s$ \ltsima $11.5$. Close to the centre, however, the density
of stars increases very rapidly, yielding less precise 
photometric measurements due
to source crowding. Another issue is the existence of selection effects: we fit
a straight line to the upper giant branch defined in a $(K_s,J-K_s)$ CMD. The
$K_s$ limit to the data will stand out clearly as a lower limit in the 
populated CMD region, but any $J$ band cut-off will be a diagonal
line on the CMD. It will bias our best-fit line if the giant branch is
crossed by this diagonal line within the magnitude
range used in the fit.
Finally, at fainter magnitudes the Bulge luminosity function rises rapidly,
increasing the contamination by Bulge stars with less extinction than the 
average, which will hamper our attempt to fit the bulk of the Bulge stellar 
population.
 
All the 3 factors just mentioned are in fact present in our data to some 
extent. Fig. 2 shows the CMDs for
two sample cells: one at $\ell=0.87^{\circ}$, ${\it b}=0.27^{\circ}$
(panels a and b), the other at $\ell=1.93^{\circ}$, ${\it b}=0.93^{\circ}$
(panels c and d). The panels on the left (right) show the data in
the range 8 $\le {\it K_s} \le $ 11.5 (9 $\le {\it K_s} \le $ 13).
The reference upper giant branch is shown in all panels, reddened by
the best-fit ($A_K$,$E(J-K)$) value found in each case.
Clearly, the extinction is underestimated for the lower latitude cell when
stars in the 9 $\le {\it K_s} \le $ 13 interval are used.
The reason for that is two-fold: there is a large number of blue
stars with $K_s$ \gtsima $11$ and there is a paucity of stars 
in the lower right
corner of the CMD in panel 2b. Both tend to shift the best fit line towards
the blue. 

The scarcity of faint red stars is due to the 2MASS $J$ band limit:
there are essentially no stars beyond the diagonal line $J = 16.5$, which 
is thus the 2MASS Point Source Catalog limiting $J$ magnitude. This empirical
limit is close to that proposed by the 2MASS collaboration (Skrutskie et al. 
1997, see also the 2MASS web site). 
The faint blue stars are probably due to contamination from Bulge stars
with less extinction and to the increased
photometric errors in these dense fields.

Note that the fit within the range 8 $\le {\it K_s} \le $ 11.5,
despite the inclusion of some residual stars bluewards
of the giant branch is not biased (panel 2a). For the cell at
$\ell=1.93^{\circ}$, ${\it b}=0.93^{\circ}$ the situation is more reassuring,
the derived $A_K$ being insensitive to the magnitude range used.
We thus conclude that a more efficient use of the CMDs in
the crowded and high extinction areas close to the Galactic plane 
is made by restricting the fit to {\it$K_s$}=11.0. 
This limit is also conveniently close to where completeness effects should
start to be significant in the 2MASS. 
For DENIS, Unavane et al. (1998) estimated the 80\% completeness level to
be at $K \simeq 10; J \simeq 13$ and the 2MASS data are about 1 mag fainter.

We conclude that in areas where extinction is around ${\it A_K} \le$ 1.5 and
photometric errors are not substantially increased by crowding
our fitting method does not suffer from any biases due to
the $J$ band cut-off. For ${\it A_K} \simeq$ 2.5, the J band detection
limit effects become dominant and the range available for the fit is
substantially shortened, rendering the extinction determination unreliable.

\begin{figure}
\begin{center}
\centerline{\psfig{file=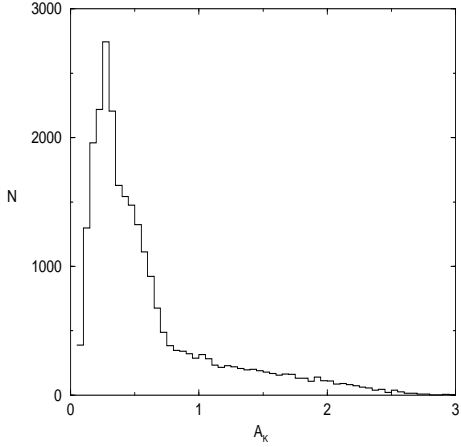,height=6cm,width=6cm,angle=0}}
\end{center}
\caption{Histogram of {\it$A_K$} extinction values of the 10$^{\circ}$ extinction map.}
\end{figure} 

\begin{figure}
\begin{center}
\centerline{\psfig{file=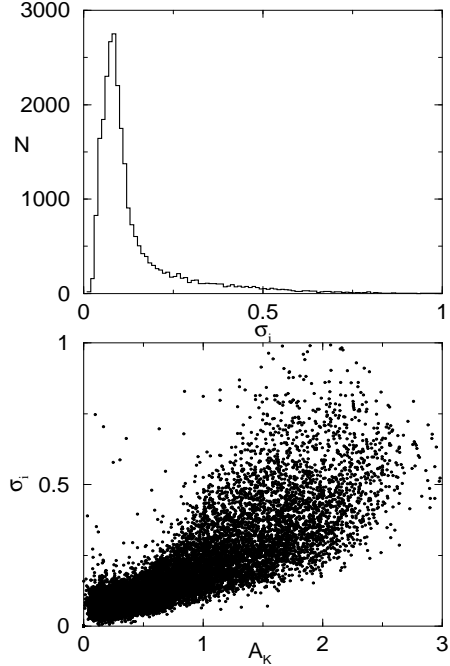,height=9cm,width=6cm,angle=0}}
\end{center}
\caption{Upper panel: histogram of internal error $\sigma_i$ values. Lower panel: Variation of $\sigma_i$ with {\it$A_K$} extinction values.}
\end{figure} 

\subsection{The {\it$A_K$} extinction map}

Fig. 3 shows the {\it$A_K$} contour map for the central 10$^{\circ} \times 
10^{\circ}$ of the Galaxy obtained by applying the method described 
in the previous section to stars down to {\it$K_s$}=11.0.
The {\it$A_K$} contours show that the regions with ${\it A_K} >$ 1.5 are 
concentrated in the area with $|{\it b}|< 1.0^{\circ}$. The isocontours with 
high values are slightly asymmetric with respect to the mid-plane and 
more extended in the southern hemisphere. They appear to be caused by the 
displaced location of the Sun from the Plane (Sect. 4). The structure 
with extinction values between {\it$A_K$}=0.5-1.0 
and angular dimension of $3^{\circ} \times 2^{\circ}$ around 
$\ell$ = 1.5$^{\circ}$  and {\it b} = 4.0$^{\circ}$ is a component of the Pipe 
Nebula, a dark nebula recently studied in CO by Onishi et al. (1999).
This nebula appears to be located at 160 pc from the Sun, as a southern
extension of the Ophiuchus dark cloud complex, and is located at the
edge of the ScoOB2 association (Onishi et al. 1999). The
detection of the Pipe Nebula in Fig. 3 suggests that nearby clouds
cannot be neglected in attempts to interpret central extinction maps.
From the 32,761 cells that cover a 12$^{\circ} \times 12^{\circ}$ central 
area of the Galaxy, we have 2MASS data for 80 \% of them. 
The quadrant $0^{\circ}<\ell<5^{\circ}$, $-5^{\circ}<{\it b}<0^{\circ}$, 
which comprises Baade's Window, is 
so far only partially released by 2MASS. 

Fig. 4 shows a histogram of {\it$A_K$} values for the cells in the 
10$^{\circ} \times 10^{\circ}$ extinction map. 
The mean extinction in the entire map is ${\it<A_K>} = 0.29$
with a standard deviation $\sigma = 0.12$ from the mean. 
63 \% of the cells fall within 2-$\sigma$ of this mean value, and 80\% 
of them have ${\it A_K} < 1.0$. The upper panel of Fig. 5 shows the 
histogram of internal errors in extinction determination. The mean 
internal error is $<\sigma_i> = 0.08$, with a standard deviation
of $0.02$ around this mean. 70\% of the cells have internal errors 
within 2 standard deviations from the mean value 
($0.04 \leq \sigma_i \leq 0.12$).
The lower panel shows the dependence of internal errors with 
{\it$A_K$}; we note that for ${\it A_K} > 1.5$ the internal errors 
increase significantly.

\begin{figure}
\begin{center}
\centerline{\psfig{file=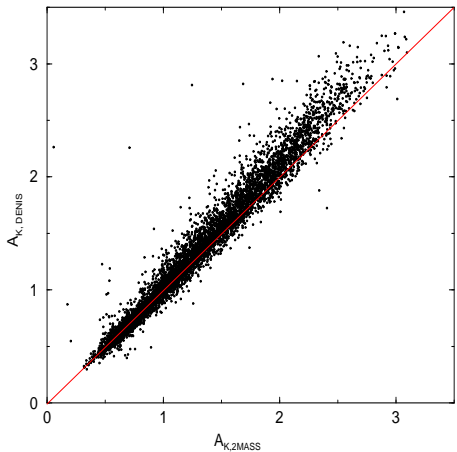,height=6cm,width=6cm,angle=0}}
\end{center}
\caption{Comparison between ${\it A_{K,2MASS}}$ and ${\it A_{K,DENIS}}$ extinction values for the region $|\ell|<5^{\circ}$ and $|{\it b}|<1.5^{\circ}$. The 
straight line represents the identity function.}
\end{figure} 

Since the present extinction map can be useful for a wide variety of Galactic 
and extragalactic studies in such central directions, 
it will be provided in electronic form in the CDS, by columns: (1) and (2) 
galactic longitude and latitude of the cell centre, (3) the {\it K} band 
extinction {\it$A_K$} and (4) the uncertainty in the {\it$A_K$} 
determination $\sigma_i$.  

\section{A 2MASS $vs.$ DENIS extinction comparison}

Schultheis et al. (1999) provided an extinction map for the inner 
Galactic Bulge (covering $|\ell|<8^{\circ}$ and  $|{\it b}|<1.5^{\circ}$) 
obtained from the {\it J} and {\it$K_s$} DENIS CMDs, together with isochrones 
from Bertelli et al. (1994), hereafter called DENIS extinction map. 
The differences between the present extinction determination 
applied to 2MASS photometry and the procedure adopted by 
Schultheis et al. (1999) are the following: (i) the latter use an 
isochrone from Bertelli et al. (1994), with metallicity {\it Z}=0.02, 
age of 10 Gyr and distance {\it d}=8 Kpc, to represent the 
dereddened ({\it K,J-K}) CMD of the Bulge fields;
(ii) Schultheis et al. adopt a transformation from {\it K} to {\it$K_s$} 
with an estimated error of 0.04 mag, in order to be able to use isochrone 
fitting; and 
(iii) the DENIS infrared photometry is limited to {\it$K_s$}=11.0,
with detection limits at $J = 16.0$ and $K_s = 13.0$ (Schultheis et al. 1999).

Unavane et al. (1998) estimated a DENIS completeness 
limiting magnitude of
{\it$K_s$}=10.0 in the inner Bulge. The resolution of the DENIS extinction 
map is also 4$^{\prime}$. 
Fig. 6 shows the comparison of the extinction values derived from the 
2MASS photometry, ${\it A_{K,2MASS}}$, with those derived 
from the DENIS photometry, ${\it A_{K,DENIS}}$, for the area in common 
between the two maps ($|\ell|<5^{\circ}$ and $|{\it b}|<1.5^{\circ}$). 
The extinction values derived from the 2MASS and DENIS photometric data 
present an excellent agreement, especially up to {\it$A_K$}=1.0. Beyond 
this limit ${\it A_{K,DENIS}}$ values are higher than ${\it A_{K,2MASS}}$ 
ones, but if we consider 
that the uncertainties in the extinction determination and photometric errors 
increase in these zones, the agreement is still significant. 
The departure from identity line in Fig. 6 could be partially due to 
differences in the filter 
profiles and zero-point calibrations adopted by the two surveys.

\begin{figure}
\begin{center}
\centerline{\psfig{file=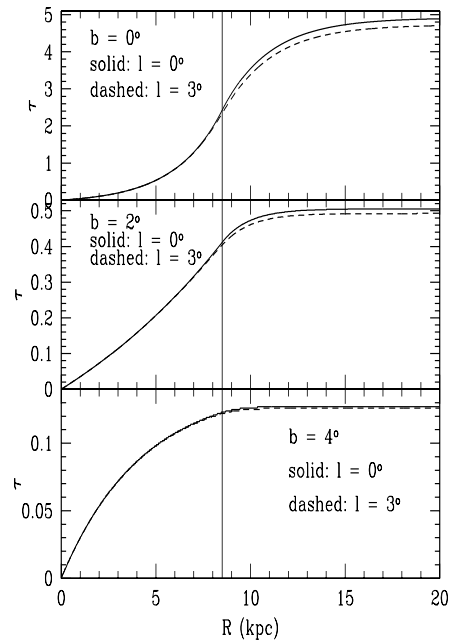,height=9cm,width=6cm,angle=0}}
\end{center}
\caption{Increase in optical depth as a function of distance from the Sun
as predicted by a simple model for the dust distribution in the Galaxy. 
The upper panel shows
directions along the disk plane, for two values of longitude $l$,
as indicated. The mid (lower) panel shows directions with
$b = 2^{\circ}~(4^{\circ})$. }
\end{figure} 

\begin{figure}
\begin{center}
\centerline{\psfig{file=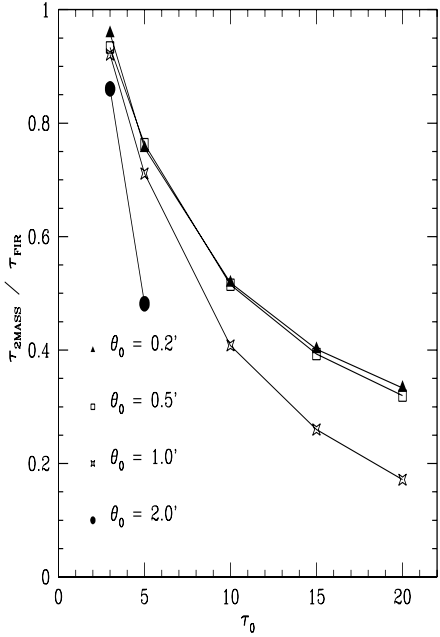,height=9cm,width=6cm,angle=0}}
\end{center}
\caption{Relative loss of sensitivity in the 2MASS extinction determination
method due to high density intervening dust clouds. The optical depths
are assumed to be averages over the entire cell of $6.1'$ in side. 
The cloud column density profile is assumed to be exponential, 
with scale length $\theta_0$ and central column density $\tau_0$.
$\tau_{FIR}$
is averaged over the entire cell but the 2MASS value is assumed to be 
restricted only to points with $\tau \leq 2$. The ratio $\tau_{2MASS} / 
\tau_{FIR}$ is plotted against $\tau_0$ for different values of $\theta_0$,
as indicated.}
\end{figure} 

\section{A 2MASS $vs.$ FIR extinction comparison}

Schlegel et al. (1998) (hereafter SFD98) 
presented an all-sky reddening map based on the 
100 $\mu$m dust emission, modeling the emission by dust grains with 
blackbody radiation at a temperature {\it T}=18.2 K. Temperature corrections 
were applied by means of the 100 and 240 $\mu$m DIRBE maps, up to $T = 21$K and
in low angular resolution ($1^{\circ}$). 
Their {\it E(B-V)$_{FIR}$} values 
correspond to the integrated dust column contribution throughout the Galaxy. 
The transformation from {\it E(B-V)$_{FIR}$} to $A_{K,FIR}$ assumed $A_K = 0.112 A_V$ and $R_V = A_V/E(B-V) = 3.1$ (Cardelli et al. 1989).
The 2MASS extinction values, on the other hand, are based on the position
of Bulge red giant stars in the CMD and, as such, 
are limited in depth by opacity effects in the employed bands, especially 
in the J band. 

In this section we compare the {\it$A_K$} extinction values 
derived from dust emission, ${\it A_{K,FIR}}$, to those from the 2MASS 
photometry, $A_{K,2MASS}$. But before we can properly compare these two
extinction maps, we must first investigate
the possible existence of systematic effects on this comparison, such
as variations in dust temperature and line of sight distribution or the
presence of high density intervening clouds.
Such effects should arise due to the
widely different observational 
signatures from the interstellar medium on which the two
extinction values are based.

\subsection{Foreground dust vs. background dust}

In order to help interpreting the relation between 
${\it A_{K,2MASS}}$ and ${\it A_{K,FIR}}$, we 
consider a simple model of dust distributed exponentially along and 
perpendicular to the Galactic plane. The optical depth 
out to some distance {\it r} from the Sun in the direction given by 
Galactic coordinates ($\ell$,{\it b}) is then: 

\begin{equation}
\tau(r,l,b) \propto \int_0^r e^{-R(r,l,b)/R_d} e^{-|Z(r,b)|/Z_d},
\end{equation}

\noindent where ${\it Z = r sin b}$ and ${\it R^2 = R_0^2 + r^2cos^2b -
2~R_0~r~cosb~cos\ell }$ are cylindrical coordinates centred on the Galaxy.
${\it R_d}$ and ${\it Z_d}$ are the dust horizontal and vertical exponential 
scales, whose values we assume to be 2.5 kpc (Robin et al. 1996 and 
Drimmel \& Spergel 2001) and 110 pc (Mendez \& van Altena 1998), respectively.
We also adopt ${\it R_0} = 8.5$ kpc for the Sun's distance to the centre of the
Galaxy.
We numerically performed the integral above for the same directions for which
we have 2MASS and DIRBE/IRAS data. For each direction, we 
assume that our model ${\it A_{K,2MASS}}$ and ${\it A_{K,FIR}}$ values are 
proportional to $\tau(R_0)$ (hereafter $\tau_{8.5}$) and 
$\tau(\infty)$ (hereafter $\tau_{\infty}$), respectively.
To model the foreground and background distribution with respect to the 
Galactic Centre, we also assume that the solar position is displaced by
${\it Z_{sun}} = 15$ pc above 
the Galactic plane (e.g. Cohen 1995); in the double exponential dust 
distribution model, the fraction of foreground-to-total dust distribution does
not depend on whether the Sun is displaced above or below the Galactic Plane.
A full data {\it vs.} models comparison is currently under way. For now
we restrict our discussion to the basic model features and their differences
relative to the data.

Fig. 7 shows $\tau$ as a function of distance from us for several
directions $(l,b)$ as predicted by our simple model. The figure clearly
shows the dependence on Galactic latitude of the 
expected contribution of dust beyond $R_0 = 8.5$ kpc. The fraction of
the total optical depth caused by dust beyond the Galactic centre varies
from 50 \% at the Galactic Plane to only a few percent for
$b = 4^{\circ}$. Fig. 7 also shows that the total optical depth itself
decreased by nearly two orders of magnitude between these two Galactic
latitudes. The dependence on $\ell$, on the other hand, is quite small.

\subsection{Resolution elements and zones of avoidance}

As we determined $A_{K,2MASS}$ in cells of 4' on a side, the resolution
of our maps is larger than the 6.1' resolution of the SFD98 extinction maps.
In order to place both maps on a similar angular resolution, we 
convolve the ${\it A_{K,2MASS}}$ extinction map with a 
$\sigma = 4.5^{\prime}$ Gaussian. 

However, there are biases that do not depend on the resolution scale of
the maps. One important issue is that the $A_{K,2MASS}$ values require
a minimum number of stars along the upper giant branch to be determined.
Regions with $A_K$ \gtsima $2$ may be obscured enough that only the
brightest stars will fall in the range $K_s \leq 11$ used to determine
$A_{K,2MASS}$ (see Sect. 2.1). If the region covers the entire cell, then no
extinction value will result. But more common will be the cases when
this zone of avoidance partially covers a cell. In that case, the cell may
still have stars enough for the giant branch fitting method to be applied
but the resulting $A_{K,2MASS}$ will be an underestimate of the true
one. Note that the $A_{K,FIR}$ values will not suffer from such bias.
We may model the latter simply as an average over the entire cell:

$$\tau_{FIR} = { {\int_{cell} \tau(\omega) d\omega} \over {\int_{cell} 
d\omega} }$$

\noindent where the integral is over the cell's solid angle and 
$\tau (\omega)$ is the varying $K$ band optical depth within the cell
 boundaries. For the 2MASS, assuming that only regions with $\tau < 2$ will 
contribute with CMD stars, we would have:

$$\tau_{2MASS} = { {\int_{< 2} \tau(\omega) d\omega} \over {\int_{cell} 
d\omega} }$$

\noindent where the 2MASS integral limit on the numerator corresponds
to the cell regions where $\tau < 2$. We may then
estimate the relative bias in the latter quantity by determining
the ratio $\tau_{2MASS} / \tau_{FIR}$ for some model for $\tau (\omega)$.
An important case of interest, especially for the lowest $b$ regions 
($|b| < 0.5^{\circ}$ (see Fig. 3), will
be that of a direction cutting through a dense dust cloud. We may consider
that the cloud will dominate the dust column density in that direction and
therefore ignore the contribution of foreground and background material.
If we also assume the cloud to have a central optical depth $\tau_0$,
circular symmetry, 
and an exponentially falling dust column density, we will have:

$$\tau (\omega) = \tau(\theta) = \tau_0~e^{-\theta / \theta_0}$$

\noindent where $\theta$ is the angle between the direction to the 
cloud centre and the
direction considered and $\theta_0$ is the cloud exponential scale.

Fig. 8 shows $\tau_{2MASS} / \tau_{FIR}$ as a function of $\tau_0$ for
several choices of $\theta_0$. For $\theta_0 = 2'$ and $\tau_0 > 5$, the
entire cell would have $\tau > 2$ and therefore $\tau_{K,2MASS}$ 
would become undetermined.
We infer from the figure that the bias on the 2MASS data 
is relatively insensitive to the cloud profile shape but it is 
clearly dependent
on profile normalization. For dust clouds with $\tau_0 < 5$ (roughly 
corresponding to a central $A_V < 55$), 
the systematic effect on $A_{K,2MASS}$ will be \ltsima $20$\%.
Note that these ratios were computed assuming the centre of the dust cloud
to coincide with the cell's centre. For the more common off-center positions, 
$A_{K,2MASS} / A_{K,FIR} \simeq \tau_{K,2MASS} / \tau_{FIR}$ will 
be larger than shown in Fig. 8. Also, in the higher extinction regions
there may possibly be more than one cloud per cell, thus increasing 
this systematic effect.

\begin{figure}
\begin{center}
\centerline{\psfig{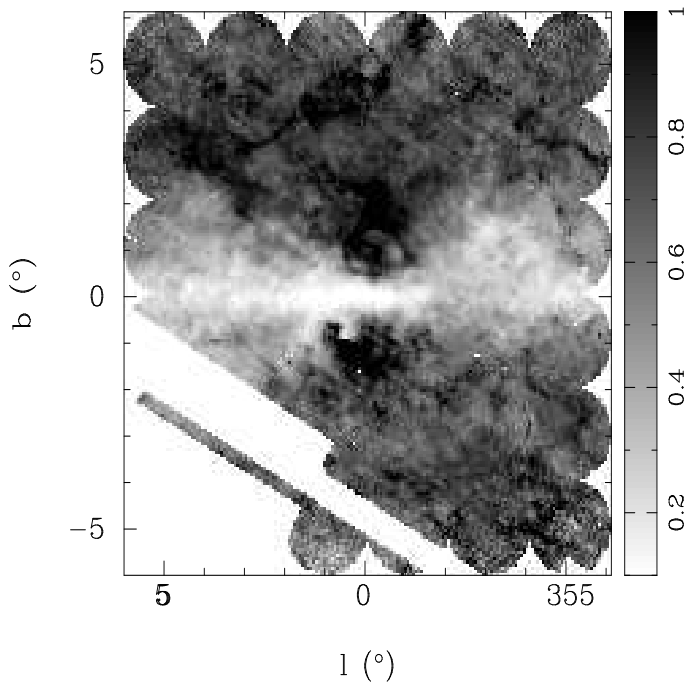}}
\end{center}
\caption{{\it$A_{K,2MASS}/A_{K,FIR}$} ratio map within 10$^{\circ} 
\times 10^{\circ}$ of the Galactic centre. Large (small) ratio values 
correspond to dark (light) areas as indicated in the greyscale bar shown
on the right.}
\end{figure}

\begin{figure}
\begin{center}
\centerline{\psfig{file=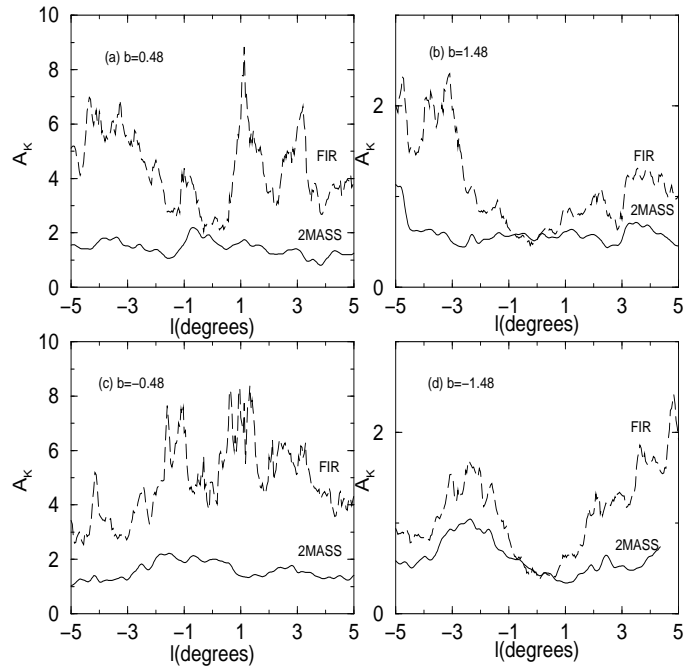,height=9cm,width=9cm,angle=0}}
\end{center}
\caption{$A_{K,2MASS}$ (solid line) and $A_{K,FIR}$ (dashed line) profiles in galactic longitude centred in the galactic latitudes:(a) $b=0.48^{\circ}$,
(b) $b=1.48^{\circ}$, (c) $b=-0.48^{\circ}$ and (d) $b=-1.48^{\circ}$.}
\end{figure}

\begin{figure}
\begin{center}
\centerline{\psfig{file=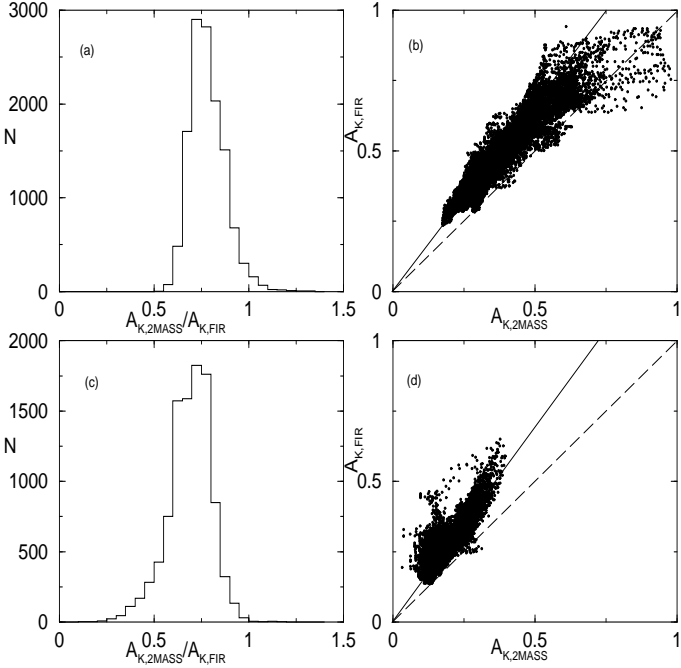,height=9cm,width=9cm,angle=0}}
\end{center}
\caption{Comparison between ${\it A_{K,2MASS}}$ and ${\it A_{K,FIR}}$ 
extinction values for cells with $5^{\circ}>{\it b}>3^{\circ}$ (upper panels) 
and $-3^{\circ}>{\it b}>-5^{\circ}$ (lower panels). Panels (a) and (c) show the
$A_{K,2MASS}/A_{K,FIR}$  histogram, whereas panels (b) and (d) show the
comparison over the entire range in {\it$A_K$}.
The dashed lines in the latter panels represent the identity function, 
whereas the solid lines have angular coefficients equal to the median
$A_{K,2MASS}/A_{K,FIR}$ value within each strip.} 
\end{figure} 

\begin{figure}
\begin{center}
\centerline{\psfig{file=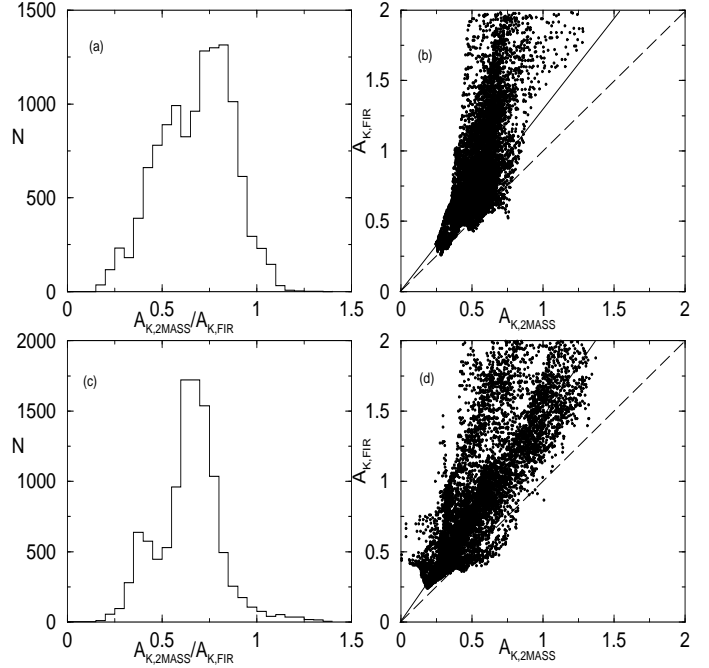,height=9cm,width=9cm,angle=0}}
\end{center}
\caption{Comparison between ${\it A_{K,2MASS}}$ and ${\it A_{K,FIR}}$ 
extinction values for cells with $3^{\circ}>{\it b}>1^{\circ}$ (upper panels) 
and $-1^{\circ}>{\it b}>-3^{\circ}$ (lower panels). Panels (a) and (c) show the
$A_{K,2MASS}/A_{K,FIR}$  histogram, whereas panels (b) and (d) show the
comparison up to $A_K= 2.0 $. 
The dashed lines in the latter panels represent the identity function, 
whereas the solid lines have angular coefficients equal to the main peak
in the corresponding $A_{K,2MASS}/A_{K,FIR}$ histograms.} 
\end{figure}

\subsection {$A_{K,2MASS}/ A_{K,FIR}$ ratio map}

Fig. 9 shows a gray scale map of the $A_{K,2MASS}/ A_{K,FIR}$ ratio, 
which provides information about the relative contribution 
of dust on the background of the Galactic centre, as background dust affects
only $A_{K,FIR}$. Since this is a ratio
map between 2 observed extinction distributions, the result is 
not dependent on the calibration details.
We note a global effect in the sense that
$A_{K,2MASS} < A_{K,FIR}$, which will be discussed later. One clearly 
sees from the ratio map the existence of a flat region 
of very low $A_{K,2MASS} / A_{K,FIR}$ ratio. In this region closest 
to the Plane, the FIR extinction values reach much
larger optical depths than the 2MASS data. Besides, the FIR extinction 
values are severely overestimated by FIR emission of dust heated at 
temperatures above those assumed in the correction based on the 100 and 
240 $\mu$m DIRBE maps.
There are also two symmetric dark regions  
close to the Galactic Centre, where $A_{K,2MASS}$ is comparable to 
$A_{K,FIR}$. These structures suggest a conical region depleted of dust
on either side of and centred at the nucleus. 
This interpretation is based on the more limited optical depth of the
2MASS extinction values, and on the fact that the 
features on the foreground of the Galactic centre are present both in the 
2MASS and FIR maps and should therefore be canceled
out. This conical structure is similar to those observed in emission
line maps from some AGNs (e.g. Storchi-Bergmann \& Bonatto 1991).
It might be related to the central low luminosity AGN in our Galaxy, with an
ionizing cone orientation nearly coincident with that of the Galaxy rotation 
axis. However, it may more readily be explained by a more extended central 
dust disc hole producing a shadow effect
on the  ionizing radiation. It would deplete dust grains within a cone, and
the UV radiation sources are more probably related to
continuous star formation in the central parts
(e.g. Arches and Quintuplet clusters presently) and/or Post-AGB stars from
the central Bulge (Binette et al. 1994).

Another way of assessing the effect of dust depletion in the cone region is
to show cuts along galactic longitude at different latitudes.
This is done in Fig. 10, where we show profiles of 
$A_{K,2MASS}$ (solid line) and $A_{K,FIR}$ (dashed line) along 
galactic longitude in positions towards the conical structure. 
We clearly see the effect of central dust depletion in 
panels (10b) and (10d), which is not expected at these galactic latitudes.

One could naively argue that $A_{K,2MASS} < A_{K,FIR}$ because
of dust emission beyond the Galactic centre, which contributes 
only to $A_{K,FIR}$. Since the relative contribution 
of background dust should be the major reason for any discrepancy
between 2MASS and SFD98 extinction
values, we will analyze them in separate for 
three regions, each one divided into a northern and a southern 
strip: (1) $|\ell|<5^{\circ}$ and 
$5^{\circ}>|{\it b}|>3^{\circ}$, (2) $|\ell|<5^{\circ}$ and 
$3^{\circ}>|{\it b}|>1^{\circ}$, and (3) $|\ell|<5^{\circ}$ and 
$|{\it b}|<0.5^{\circ}$.  
This division allows assessment of 
the correlation between ${\it A_{K,FIR}}$ and ${\it A_{K,2MASS}}$
in regions with varying contributions by the
background dust. We hope therefore to disentangle the effect on
${\it A_{K,FIR}}$ by dust on the far side of the Galaxy
from other factors that may lead to discrepancies in the 
${\it A_{K,FIR}}~vs.$ ${\it A_{K,2MASS}}$ relation. 

\subsubsection {Strips $5^{\circ}>|{\it b}|>3^{\circ}$}

Fig. 11 shows the ${\it A_{K,FIR}}$ $vs.$ ${\it A_{K,2MASS}}$ relation
in the region $5^{\circ}>|{\it b}|>3^{\circ}$. The upper panels correspond 
to cells in the northern Galactic strip ($|\ell|<5^{\circ}$, 
$5^{\circ}>{\it b}>3^{\circ}$), whereas the lower panels correspond to the 
southern ($|\ell|<5^{\circ}$, $-3^{\circ}>{\it b}>-5^{\circ}$) strip. 
The $A_{K,2MASS}/ A_{K,FIR}$ ratio histograms in panels (11a) and (11c) show 
that the relative extinction values have peaks at 0.75 (75 \%) and 0.72 
(72 \%) in the northern and southern strips, respectively. These 
well-defined peaks imply a good linear correlation between the two 
extinction values, as confirmed in panels (11b) and (11d). 
For the northern and southern strips we have the correlations 
${\it A_{K,FIR}}=1.33 {\it A_{K,2MASS}}$ and 
${\it A_{K,FIR}}=1.39 {\it A_{K,2MASS}}$, respectively, 
where the angular coefficients were derived from the median 
$A_{K,2MASS}/ A_{K,FIR}$ ratio values.
In this region, the model predictions indicate that, for both northern and 
southern strips, the ratio of extinction on foreground of the Galactic 
to total extinction should be 96\%, independent of the Sun's displacement from
the disc mid-plane. From the median $A_{K,2MASS}/ A_{K,FIR}$ values 
we infer that a typical value for this ratio is $\approx$ 73\%.
Therefore this discrepancy between model and observed extinction ratio 
suggests that the contribution from background dust
cannot explain the observed $A_{K,2MASS}/ A_{K,FIR}$ ratio.
  
Arce \& Goodman (1999) also found a 
linear relation between the dust emission extinction and that derived from 
the stellar content in the Taurus Dark Cloud, with a slope that varies  
in the range 1.3-1.5. They attributed this difference 
to the dust column density $vs.$ reddening calibration from 
Schlegel et al. (1998), which may yield an 
overestimated ${\it A_V}>$0.5. Note that the slopes in our two strips 
(with 2 $> A_V >$ 10) are similar to those of Arce \& Goodman (1999), 
which corroborates the idea of a $A_{K,FIR}$ calibration effect.

We point out that assuming a significantly lower 
value for $R_V$ would compress the $A_{K,FIR}$ scale, decreasing
the differences between $A_{K,FIR}$ and $A_{K,2MASS}$ in Fig. 11.
However, observational constraints on $R_V$ do not support this possibility.
Indeed, Gould  et al. (2001) estimated $R_{VI}$ = $A_V /E(V-I)$ 
= 2.4 ($R_V \approx$ 3.0), from $VIK$ colours of 146 Baade's Window G and K 
giants. Stanek (1996) obtained $R_{VI}$ = 2.5$\pm 0.1$ ($R_V = 3.1$), using 
Baade's Window red clump giants. For stellar fields and reddened metal-rich 
globular clusters throughout the Bulge, values of $R_V = 3.5-3.6$ have been 
employed (Terndrup 1988, Barbuy et al. 1998). We conclude that typical $R_V$ 
values in Bulge directions cannot be significantly lower than $R_V = 3.1$.
 
Assuming that the discrepancy between model and observed extinction ratio in 
these strips is entirely due to a calibration problem, we estimate a 
calibration factor of 76\% to be applied to the $A_{K,FIR}$ values.

Panels (11b) and (11d) also reveal a strong asymmetry between the northern 
and southern strips.
{\it$A_K$} values reach up to {\it$A_K$}=1.0 in the north, with 
most cells in the range $0.2< {\it A_{K,2MASS}}<0.7$. In the south,
there are few cells with ${\it A_{K,2MASS}}>0.4$.
In Sect 4.4, we discuss the north-south asymmetry in more detail.

\begin{figure}
\begin{center}
\centerline{\psfig{file=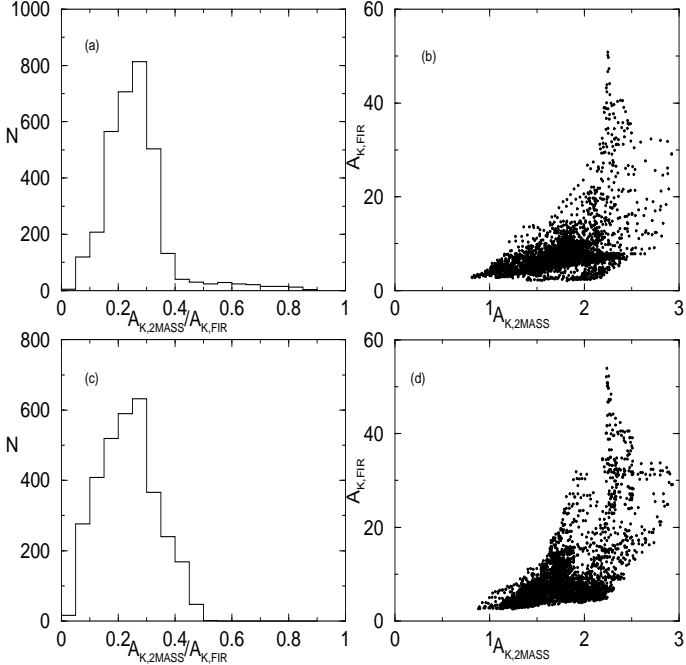,height=9cm,width=9cm,angle=0}}
\end{center}
\caption{Comparison between ${\it A_{K,2MASS}}$ and ${\it A_{K,FIR}}$ 
extinction values for cells with $0^{\circ}<{\it b}<0.5^{\circ}$ (upper panels) 
and $0^{\circ}>{\it b}>-0.5^{\circ}$ (lower panels). Panels (a) and (c) show the
$A_{K,2MASS}/A_{K,FIR}$  histogram, whereas panels (b) and (d) show the
comparison over the entire range in {\it$A_K$}.}
\end{figure}  

\subsubsection {Strips $3^{\circ}>|{\it b}|>1^{\circ}$}

As we consider regions closer to the Galactic plane, several predictions
can be made regarding the extinction values derived from the 2MASS and 
DIRBE/IRAS data. First, one obviously expects a general increase in both
${\it A_{K,2MASS}}$ and ${\it A_{K,FIR}}$ values. Further, the relation 
between the two should increasingly depart from the identity line, as 
the contribution of background dust is enhanced. Finally, line-of-sight 
variations should also grow in amplitude, as more individual dust clouds, 
with variable dust densities and temperatures, are expected to lie
along low latitude directions. 

All these predictions are confirmed by inspection of Fig. 12, which
shows the comparison between ${\it A_{K,2MASS}}$ and ${\it A_{K,FIR}}$ 
for the region $3^{\circ}>{\it b}>1^{\circ}$. The $A_{K,2MASS}/A_{K,FIR}$ 
histograms in panels (12a) and (12c) are dominated by double peak 
distributions, 
at $A_{K,2MASS}/A_{K,FIR}$ = 0.75 and 0.60 for the northern strip, and at 
$A_{K,2MASS}/A_{K,FIR}$ = 0.65 and 0.35 for the southern strip.
As a consequence, the ${\it A_{K,FIR}}~vs.$ $A_{K,2MASS}$ relation given in 
panels (12b) and (12d) is more complex and not well described by a linear fit.
Notice that the double peak nature is clearly seen in panel (12d) as two
distinct branches in the ${\it A_{K,FIR}}~vs.$ $A_{K,2MASS}$ relation. 
The solid lines shown in these panels have angular coefficients 
derived from the main peak of the $A_{K,2MASS}/A_{K,FIR}$ histograms in 
Panels (12a) and (12c).

For this region, the model predictions give a foreground 
to total extinction ratio of 81\%. If we artificially introduce in the
model the calibration factor obtained for $A_{K,FIR}$ in the previous section,
this ratio is reduced to 62\%. This is in agreement with the 
typical $A_{K,2MASS}/A_{K,FIR}$ value found in the two strips. 
The main peak in the northern strip (with a ratio of 75\%)
indicates an observed excess of foreground dust with respect to the 
dust exponential model. The secondary peak in the north matches quite well the
calibrated model values. The same applies to the main peak in the southern 
strip, whose ratio is 60\%. The secondary peak in the southern strip
(at $A_{K,2MASS}/A_{K,FIR}$ = 35\%) suggests that the observed $A_{K,FIR}$ is 
excessively high as compared to the expectations of our simple model.
This indicates lines of sight more strongly affected by dense dust clouds
and temperature effects caused by dust heated over the DIRBE/IRAS temperature 
correction. 

\subsubsection {Strips $|{\it b}|<0.5^{\circ}$}

Very close to the Galactic Plane, the infrared emission by hot dust 
near the Centre, across spiral arms or close to
star-forming regions, may increase substantially the estimate
of ${\it A_{K,FIR}}$. Therefore, this latter is expected to deviate 
significantly
from ${\it A_{K,2MASS}}$ and to display strong variations due to scatter in
line-of-sight dust. Fig. 13 shows ${\it A_{K,FIR}}~vs.$ 
${\it A_{K,2MASS}}$ for the $|{\it b}|<0.5^{\circ}$ strips. The 
$A_{K,2MASS}/A_{K,FIR}$ histograms in panels (13a) and (13c) 
show peaks at $A_{K,2MASS}/A_{K,FIR}$ = 0.27 (27\%) for both northern and 
southern strips. The models predict a relative extinction of 54\% in the same 
region; introducing the $A_{K,FIR}$ calibration effect found previously, 
this value changes to 41\%. Therefore, the observed extinction ratios
are systematically smaller than the model expectations.
As discussed in Sect. 4.2, this is just what one expects from the existence of
non-exponential dust distribution (dense dust clouds) in the foreground
and background of the Galactic Centre, and the overestimation of 
the $A_{K,FIR}$ due to heated dust. Furthermore, as discussed in the end
of Sect. 2.1, $A_{K,2MASS}$ becomes undetermined 
for $A_K >$ 2.5 due to the limit in the J band optical depth. 
In fact, this cut-off in the $A_{K,2MASS}$ values is clearly seen in
panels (13b) and (13d), as $A_{K,2MASS}$ saturates for rising
$A_{K,FIR}$ values.

\subsection{North-South asymmetry}

A better way of assessing any north-south discrepancy in the
$A_{K,2MASS}$ and $A_{K,FIR}$ data is to directly compare 
cells with the same value of $\ell$ but with opposite signs of {\it b}. 
This method has the additional advantage that the comparison is not affected
by the incomplete spatial coverage in the southern strip. 

\begin{figure}
\begin{center}
\centerline{\psfig{file=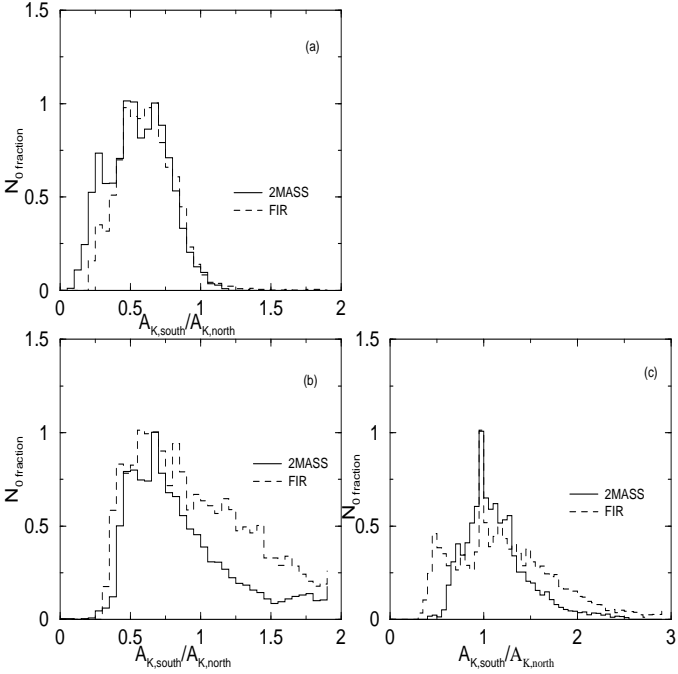,height=9cm,width=9cm,angle=0}}
\end{center}
\caption{$A_{K,2MASS}(south)/A_{K,2MASS}(north)$ (solid line) and $A_{K,FIR}(south)/A_{K,FIR}(north)$ (dashed line) for cells at $3^{\circ}<{\it |b|}<5^{\circ}$ (panel a),
 $1^{\circ}<{\it |b|}<3^{\circ}$ (panel b), and $0^{\circ}<{\it |b|}<0.5^{\circ}$ (panel c).
}
\end{figure} 

Fig. 14 shows histograms of the ratio of extinction values in 
mirror cells relative to the disc plane. Panels (a), (b) and (c) present 
extinction ratios for mirror cells in the strips 
$3^{\circ}<{\it |b|}<5^{\circ}$, $1^{\circ}<{\it |b|}<3^{\circ}$, and 
$0^{\circ}<{\it |b|}<0.5^{\circ}$, respectively.
The solid histograms in each panel show the distribution
of $A_{K,2MASS}(south)/A_{K,2MASS}(north)$ ratios, while the dashed histograms
correspond to the $A_{K,FIR}(south)/A_{K,FIR}(north)$ distribution.

All histograms are normalized by the main peak in counts, 
allowing a better comparison. In the region 
$3^{\circ}<{\it |b|}<5^{\circ}$, 
we have two very similar distributions, with peaks around $\approx$ 0.6.
The general agreement indicates that both extinction measurements detect 
the same asymmetry, with the extinction being 60\% less in the south than 
in the north. The strong similarity confirms that, in these particular
strips, the extinction measured in the FIR is dominated by dust on the 
foreground of the stars that were used to measure ${\it A_{K,2MASS}}$. 

In the region $1^{\circ}<{\it |b|}<3^{\circ}$, the observed asymmetry 
in the FIR and 2MASS data is again very similar in the range 
$0 < A_{K,south}/A_{K,north} < 0.7$, displaying a large peak centred around 
0.6 (60\%).
However, the FIR ratios are systematically larger than the 2MASS one
in the $A_{K,south}/A_{K,north} > 0.7$ domain. This indicates a 
systematic trend in the sense that
southern cells suffer a more significant contribution by background dust
than their corresponding mirror cells in the north.

In the region $|b|<0.5^{\circ}$, the north-south asymmetry
is substantially reduced and, interestingly enough, most cases now
have $A_{K,south}/A_{K,north} \approx$ 1.0. This applies to both 
FIR and 2MASS data,
the main difference between the two datasets being the larger dispersion
in the south/north values in the former relative to the later.

Interpreting this north-south asymmetry is not an easy task. The similarity
of the histograms in panel (14a) plus our model predictions suggest that
extinction in the $3^{\circ}<{\it |b|}<5^{\circ}$ strips is dominated
by foreground structure in the form of extended dust clouds (see also
map in Fig. 3 and the concluding section). At lower latitudes, the extinction
values increase and become dominated by diffuse dust distribution or by
dense clouds located close to the Galactic centre. In these cases,
the shape and position of the $A_{K,south}/A_{K,north}$ histograms 
in both datasets should reflect the global properties of
the dust distribution rather than the contribution of any individual 
dust cloud. As the FIR data reach deeper into the dust columns than the
2MASS, the transition from the foreground dominated regime to one
dominated by the global distribution starts earlier in the former dataset; this
effect is the likely cause of the difference in the histograms in
panel (14b). 

In the $|b|<0.5^{\circ}$ strip, the $A_{K,2MASS}$ values 
are strongly limited by optical depth and largely unreliable (Sect. 2.1), 
while the $A_{K,FIR}$ values are suffer from the effects 
of dust heated beyond 21K. Despite these uncertainties the dust distribution 
in this region is reasonably symmetric.
There is a small trend in the sense of $A_{K,south}/A_{K,north} > 1$,
especially in the FIR data. This trend
is intriguing. The fact that most cells have larger $A_{K,FIR}$ values 
at southern latitudes than at northern ones may be telling us something
about the geometry of the dust distribution relative to the Sun's position.

In order to take into account a possible displacement by the Sun relatively
to the disk mid-plane, we again use our model presented in Sect. 4.1. 
Considering the Sun displaced by 5, 15 and 25pc
above the Plane, we obtain typical south/north ratios in
the $\tau_{8.5}$ or $\tau_{\infty}$ values
of 1.08, 1.27 and 1.50, respectively. 
However, the model does not incorporate any asymmetry caused 
by foreground dust clouds, which seems to be the dominant effect generating
the asymmetry in the 2MASS and FIR data. It is thus impossible
at the present to quantify the Sun's displacement from the disc based on
the data. We should point out that
Unavane et al. (1998), using DENIS data, also concluded that the asymmetry 
in the inner Bulge extinction is dominated by foreground dust clouds.

\section{Concluding remarks}

We built the {\it$A_K$} extinction map towards the central $10^{\circ} 
\times 10^{\circ}$ of the Galaxy using the 2MASS Point Source Catalog. We 
extracted J and K$_s$ 
magnitudes for about 6 million stars in the range $8.0 \le {\it K_s} \le 13.0$.
The adopted map resolution is 4$^{\prime} \times 4^{\prime}$. It 
was possible to obtain extinction values for $\approx$80 \% of the 32,761 cells
defined in the area, where 2MASS data were currently available and a Bulge
giant branch was distinct enough. The extinction affecting
the bulk of the Bulge stellar population was determined by matching the 
upper giant branch found in the {\it$K_s$}, ({\it J-K$_s$}) colour magnitude 
diagram to the reference upper giant branch built using de-reddened 
Bulge fields. 
The extinction values vary from {\it$A_K$}=0.05 in the edges of the map up 
to {\it$A_K$}=3.2 close to the Galactic centre. The mean extinction found is
${\it<A_K>}=0.29$ with a dispersion $\sigma = 0.12$; 
63 \% of the cells are within 2-$\sigma$ of the mean. 

We compared our 2MASS extinction map to that of
Schultheis et al. (1999) in the region $|\ell| <5^{\circ}$ and 
$|{\it b}|<1.5^{\circ}$, which is common to both studies.
Schultheis et al. extinction map is based on DENIS photometry. 
We find an excellent agreement between the two extinction determinations,
especially up to ${\it A_K}=1.0$. Beyond this limit the values derived from 
the DENIS data are systematically larger. This small discrepancy
in large extinction regions is not unexpected, 
considering the photometric errors, incompleteness effects, and 
uncertainties in extinction determination. 

We also compared the present extinction map to that of Schlegel et al. (1998),
which is based on dust emission in the far infrared 
detected by the DIRBE/IRAS instruments.
As the data from the latter are affected by the entire dust column, with no
depth limit, the comparison was made separately for regions of 
decreasing Galactic latitude {\it b}, which supposedly correspond to 
increasing contribution
by dust located on the background of the Galactic Centre. 
The background dust contribution was estimated by means of a double 
exponential dust distribution model.
Some expected systematic biases, besides that caused by dust on
the far side of the Galaxy, were also assessed and quantified.

In general, the extinction values derived from dust emission are 
higher than 
those from 2MASS, mainly close to the Galactic Plane and Centre. We detected 
two unexpected regions symmetric and close
to the Galactic Centre where the two extinction estimates are of 
the same order. The lack of background dust in these
low latitude regions could be explained by a process of dust grain 
destruction by UV emission from sources 
associated with continuous star formation and/or Post-AGB stars in the 
central parts of the Galaxy.

For the cells in the region $3^{\circ}<|{\it b}|<5^{\circ}$, we observe
a clear and roughly linear correlation between the {\it$A_K$} values 
from 2MASS and dust emission. 
We also confirm, as was done in Paper I, that the {\it$A_K$} values 
from 2MASS data are in general 73\% smaller than those derived from 
dust emission. 
Since in this region the background dust contribution is less than 5 \%, 
the differences between these two quantities should be smaller than observed.
This discrepancy is also verified by Arce \& Goodman (1999) in the Taurus Dark 
Cloud. It is probably due to systematic effects in the
dust column density {\it vs.} reddening calibration from 
Schlegel et al. (1998), yielding an 
overestimate of extinction in moderate to high extinction regions.
We estimate a calibration correction
factor of 76\% for the FIR extinction values.

For the intermediate $1^{\circ}<|{\it b}|<3^{\circ}$ region, the relation 
between DIRBE/IRAS and 2MASS extinction values departs more significantly
from the identity line, as expected due to the larger contribution by 
background dust. In this region, the typical $A_{K,2MASS}/A_{K,FIR}$
ratio is 65\% and could
be explained by background dust contribution and the calibration factor
affecting the FIR data. 
An enhancement in the foreground dust with respect to 
the dust model is observed in many cells in the northern strip. 
In the southern
strip, several cells have $A_{K,2MASS}/A_{K,FIR}$ smaller than expected,
probably due to dense dust clouds and temperature variations, currently not 
incorporated into our model.

For the regions very close to the Galactic Plane ($|{\it b}| 
< 0.5^{\circ}$), we have a typical value for the $A_{K,2MASS}/A_{K,FIR}$ 
ratio of 27\%. Even considering the background dust contribution and 
the calibration factor, this ratio is still smaller than that predicted by 
our simple model for the dust distribution. This fact is probably due to the 
overestimation of the $A_{K,FIR}$ values by heated dust above that 
obtained from DIRBE temperature maps. Another possible
contribution to this difference is the existence of systematic effects
on the $A_{K,2MASS}$ values in high extinction regions ($A_{K,2MASS} > 2.5$),
where the 2MASS extinction should be significantly underestimated or 
even unreliable.
 
A systematic asymmetry in the {\it$A_K$} values relative to the plane of the
Galaxy {\it at $1^{\circ}<|{\it b}|<5^{\circ}$} is observed both in the 
2MASS and DIRBE/IRAS data. The behaviour
and amplitude of this asymmetry with position on the sky suggest that the
dominant role in creating this north-south asymmetry is a more effective 
presence of foreground dust 
clouds in the northern Galactic strips, such as the Pipe Nebula (Sect. 2.2).
A possible explanation is stellar winds and supernovae 
from nearby OB stellar associations producing dust cloud shells (Bhatt 2000).
The nearby clouds projected towards the central parts
of the Galaxy at positive latitudes belong to the Ophiuchus dust complex.
They are probably related to the association ScoOB2, which is
at a distance of 145 pc from the Sun (Bhatt 2000; Onishi et al. 1999).
ScoOB2, in turn, belongs to Upper Scorpius, which is 
the easternmost part of the  Sco-Cen Association, as 
studied by means of Hipparcos (de Zeeuw et al. 1999).

In all regions, significant substructure in the 
${\it A_{K,2MASS}} vs.$ ${\it A_{K,FIR}}$ relation is seen, with loops and arms
stretching out from the main relation. These structures are probably caused
by intervening dust clouds, with different temperatures and densities for
different lines of sight. One extremely interesting perspective is to 
model the dust distribution within the Galaxy, trying to
reproduce as close as possible the details of the {\it$A_K$} maps currently 
available. This effort demands models that incorporate, on top of
a smooth dust distribution, the effects of
individual dust clouds, spiral arms, molecular rings and other structure,
possibly with variable density contrasts and temperatures.
This effort is currently under way for the central region of the Galaxy.

\section*{Acknowledgements}
This publication makes use of data products from the Two Micron All Sky 
Survey, which is a joint project of the University of Massachusetts and 
the Infrared Processing and Analysis Center/California Institute of 
Technology, funded by the National Aeronautics and Space Administration 
and the National Science Foundation. We also use the electronic form
of the extinction maps provided by Schultheis et al. (1999) and
Schlegel et al. (1998). We thank the anonymous referee 
for his/her interesting comments and suggestions.
We acknowledge support from the 
Brazilian institutions FAPESP and CNPq. CMD acknowledges FAPESP for a 
post-doc fellowship (proc. 00/11864-6).


\begin{thebibliography}{}

\bibitem[]{} Arce H.G., Goodman A.A. 1999, ApJ, 512, L135
\bibitem[]{} Baade W. 1963, Evolution of stars and galaxies, Harvard University Press, Cambridge, Mass., p. 277
\bibitem[]{} Barbuy B., Bica E., Ortolani S. 1998, A\&A, 333, 117
\bibitem[]{} Barbuy B., Ortolani S., Bica E., Desidera S. 1999, A\&A, 348, 783
\bibitem[]{} Bertelli G., Bressan A., Chiosi C., et al. 1994, A\&AS, 106, 275
\bibitem[]{} Binette, L., Magris, C. G., Stasinska, G., Bruzual, A. G., 1994, A\&A, 292,
13
\bibitem[]{} Bhatt H.C. 2000 A\&A, 362, 715
\bibitem[]{} Cardelli J.A., Clayton G.C., Mathis J.S. 1989, ApJ, 345, 245
\bibitem[]{} Catchpole R.M., Whitelock P.A., Glass I.S. 1990, MNRAS, 247, 479
\bibitem[]{} Cohen M. 1995, ApJ, 444, 874
\bibitem[]{} de Zeeuw P.T., Hoogerwerfet R., de Bruijne J.H.J., et al. 1999, AJ, 117, 354
\bibitem[]{} Drimmel R., Spergel D.N. 2001, ApJ, 556, 181
\bibitem[]{} Dutra C.M., Bica E. 2000, A\&A, 359, 347
\bibitem[]{} Dutra C.M., Santiago B.X., Bica E. 2002, A\&A, 381, 219
\bibitem[]{} Epchtein N., de Batz B., Capoani L. et al. 1997, The Messenger, 87, 27
\bibitem[]{} Frogel J.A., Tiede G.P., Kuchinski L.E. 1999, AJ, 117, 2296 (FTK99)
\bibitem[]{} Gould A., Stutz A., Frogel J.A. ApJ, 547, 590
\bibitem[]{} Hammersley P.L., Garzon F., Mahoney T., Calbet X. 1995, MNRAS, 273, 206
\bibitem[]{} Humphreys R.M., Larsen J.A. 1995, AJ, 110, 2183
\bibitem[]{} Mendez R.A., van Altena W.F. 1998, A\&A, 330, 910
\bibitem[]{} Onishi T., et al. 1999, PASJ, 51, 871
\bibitem[]{} Ram\'\i rez S.V., Stephens A.W., Frogel J.A., DePoy D.L. 2000, AJ, 120, 833
\bibitem[]{} Robin A.C., Haywood M., Cr\'ez\'e M., et al. 1996, 305, 125
\bibitem[]{} Schlegel D.J., Finkbeiner D.P., Davis M. 1998, ApJ,500, 525 (SFD98)
\bibitem[]{} Schultheis M., Ganesh S., Simon G., et al. 1999, A\&A, 349, L69
\bibitem[]{} Skrutskie M., Schneider S.E., Stiening R., et al. 1997, in ``The Impact of Large Scale Near-IR Sky Surveys'', ed. Garzon et al., (Dordrecht:Kluwer), p. 25
\bibitem[]{} Stanek K.Z. 1996, ApJ, 460, L37
\bibitem[]{} Storchi-Bergmann, T., Bonatto, Ch. J., 1991, MNRAS, 250, 138
\bibitem[]{} Tiede G.P., Frogel J.A., Terndrup D.M. 1995, AJ, 110, 2788
\bibitem[]{} Unavane M., Gilmore G., Epchtein N., et al. 1998, MNRAS, 295, 119

\end{thebibliography}
\end{document}